\documentclass[9pt,twocolumn,twoside]{pnas-new}
\templatetype{pnasresearcharticle}

\usepackage{graphicx,xcolor,siunitx,amssymb,amsmath,hyperref}
\usepackage[version=4]{mhchem}
\usepackage[normalem]{ulem}
\DeclareSIUnit\angstrom{\text{Å}}
\DeclareMathOperator{\tr}{tr}

\newcommand{\figsref}[1]{Figs.\,\ref{#1}}

\newcommand{\figrefp}[2]{Fig.\,\ref{#1}{\textit{#2}}}

\newcommand{\subfig}[1]{\textit{#1}}

\newcommand{\figsrefp}[2]{Figs.\,\ref{#1}{\textit{#2}}}

\newcommand{\sfigref}[1]{\textit{SI Appendix}, Fig.\,S{#1}}

\newcommand{\smovref}[1]{Video S{#1}}
\newcommand{\smovsref}[2]{Videos S{#1} and S{#2}}

\newcommand{\matref}{\hyperref[matmethods]{\textit{Materials and Methods}}}

\renewcommand{\eqref}[1]{Eq.\,\textbf{\ref{#1}}}

\begin{document}

\title{Filled Colloidal Gel Rheology: Strengthening, Stiffening, and Tunability}

\author[a,b,1]{Yujie Jiang}
\author[a,c]{Yang Cui}
\author[a,c]{Yankai Li}
\author[a,d]{Zhiwei Liu}
\author[c,1]{Christopher Ness}
\author[a,d,e,1]{Ryohei Seto}

\affil[a]{Wenzhou Key Laboratory of Biomaterials and Engineering, Wenzhou Institute, University of Chinese Academy of Sciences, Wenzhou, Zhejiang 325000, China}
\affil[b]{School of Physical Sciences, University of Chinese Academy of Sciences, Beijing 100049, China}
\affil[c]{School of Engineering, The University of Edinburgh, King's Buildings, Edinburgh EH9 3FG, UK}
\affil[d]{Oujiang Laboratory (Zhejiang Lab for Regenerative Medicine, Vision and Brain Health), Wenzhou, Zhejiang 325000, China}
\affil[e]{Graduate School of Information Science, University of Hyogo, Kobe, Hyogo 650-0047, Japan}

\leadauthor{Yujie Jiang}

\significancestatement{
Solid filling is widely used for reinforcement in composite materials.
The technique is well-established in polymeric nanocomposites, but less understood in the emerging field of filled colloidal gels, relevant to development of, e.g., novel battery electrolytes.
We report the counter-intuitive experimental finding that filling colloidal gels with grains non-monotonically reinforces their mechanical properties. 
Addition of filler increases the yield strength of the gel, whereas the stiffness increases up to a point before decreasing thereafter.
Simulations reveal the underlying mechanism, which has implications for gel formulation and leads us to propose a novel `filler removal protocol,' enabling individual control over gel strength and stiffness.
Our work thus opens new avenues for characterizing, designing, and tuning the mechanics of soft materials.
}

\authorcontributions{Y.J., C.N., and R.S.\ designed research; Y.J., Y.C., and Y.L.\ performed research; Y.J., Y.C., Y.L., and Z.L.\ analyzed data; Y.J., C.N., and R.S.\ wrote the paper.}
\authordeclaration{The authors declare no competing interest.}
\correspondingauthor{\textsuperscript{1}To whom correspondence should be addressed. E-mail: yujiejiang\_1994@outlook.com; chris.ness@ed.ac.uk; seto@ucas.ac.cn}

\keywords{colloidal gels $|$ solid filling $|$ rheology $|$ tunability}

\begin{abstract} 
Filler-induced strengthening is ubiquitous in materials science and is particularly well-established in polymeric nanocomposites.
Despite having similar constituents, colloidal gels with solid filling exhibit distinct rheology, which is of practical interest to industry (e.g., lithium-ion batteries) yet remains poorly understood.
We show, using experiments and simulations, that filling monotonically enhances the yield stress (i.e., strength) of colloidal gels while the elastic modulus (i.e., stiffness) first increases and then decreases.
The latter softening effect results from a frustrated gel matrix at dense filling, evidenced by a growing inter-phase pressure.
This structural frustration is, however, not detrimental to yielding resistance.
Instead, fillers offer additional mechanical support to the gel backbone via percolating force chains, decreasing the yield strain at the same time.
We develop a mechanistic picture of this phenomenology that leads us to a novel `filler-removal protocol,' making individual control over the strength and brittleness of a composite gel possible.
\end{abstract}

\dates{This manuscript was compiled on \today}
\doi{\url{www.pnas.org/cgi/doi/10.1073/pnas.XXXXXXXXXX}}

\maketitle
\thispagestyle{firststyle}
\ifthenelse{\boolean{shortarticle}}{\ifthenelse{\boolean{singlecolumn}}{\abscontentformatted}{\abscontent}}{}

\firstpage[5]{4}


\dropcap{A}ttractive colloids self-assemble into percolating, porous gel networks, leading to soft solids with finite yield stress\,\cite{JPhysCondensMatter.33.453022,RevModPhys.89.035005,PhysRevLett.106.248303,NatPhys.19.1178,PNAS.118.e2022339118}.
These colloidal gels constitute an important material class in industries ranging from consumer products to coatings and biotechnology\,\cite{Nature.505.382-385,FaradayDiscuss.158.9-35,JohnWiley.9781119220510}.
Hence, understanding and controlling their rheology is of broad fundamental interest and practical importance.
The intrinsic dynamical arrest associated with colloidal gelation\,\cite{PhysRevLett.95.238302,NatMater.7.556-561,PNAS.109.16029-16034,NatPhys.2024,Nature.453.499-503} naturally frustrates mechanical tunability, though several recent works seek to achieve this via non-equilibrium protocols comprising external flow\,\cite{SoftMatter.11.4640-4648}, acoustic vibration\,\cite{PhysRevX.10.011028}, and active doping\,\cite{PhysRevLett.131.018301,ACSNano.13.560-572}. 
Nonetheless, the exploration of more efficient rheological control strategies is still in progress.


Solid filling is a widely-used technique for reinforcement, such as in polymer nanocomposites\,\cite{MaterTodayProc.45.2536-2539}, filled resins\,\cite{JMaterSci.19.473-486}, and concretes\,\cite{MaterStruct.37.74-81}.
The size disparity between the constituents (i.e., matrix $\sim \si{\angstrom}$ vs.\ filler $\gtrsim \si{\nano\meter}$) generally enables these materials to be characterized by application of standard approaches in continuum mechanics\,\cite{SoftMatter.10.13-38,JRheol.52.287-313}.
In particle-filled colloidal gels, these lengthscales are comparable ($\sim \si{\micro\meter}$), and their interplay generates novelty not describable by conventional models.
For example, adding non-Brownian grains to a colloidal gel leads to a mechanorheological material in which external flow triggers a unique bistability\,\cite{PhysRevLett.128.248002}. 
Meanwhile, filling non-trivially distorts the gelation diagram boundaries\,\cite{SoftMatter.19.1342-1347}, governed in part by competing lengthscales that emerge\,\cite{NatCommun.14.2773}.


As the addition of fillers reduces the volume available for colloids, the simplest characterization of a filled gel is by its effective volume fraction $\phi_{\mathrm{eff}}$, subtracting the filler volume from the total as:
\begin{equation}
\phi_{\mathrm{eff}} \equiv \frac{V_{\mathrm{gel}}}{V_{\mathrm{total}}-V_{\mathrm{filler}}} = \frac{\phi_{\mathrm{g}}}{1-\phi_{\mathrm{f}}},
\label{phi_eff}
\end{equation}
where $\phi_{\mathrm{g}}$ and $\phi_{\mathrm{f}}$ refer to the absolute volume fractions of gel colloids and filler particles, respectively.
Hence, a filled gel is more concentrated than an unfilled gel at the same $\phi_{\mathrm{g}}$.
Filler-induced strengthening, reported in various colloidal gels\,\cite{SoftMatter.18.2842-2850,Langmuir.34.3021-3029}, is qualitatively described in part by the increase in $\phi_{\mathrm{eff}}$, yet quantitatively this measure fails to capture the filled gel properties\,\cite{NatCommun.14.2773,SoftMatter.11.8818-8826}.
With increasing attention being paid to composite materials, as well as their utility in technological applications (e.g., lithium batteries\,\cite{AdvEnergyMater.1.511-516}), solid filling is considered a promising approach to tuning gels efficiently\,\cite{PhysRevE.77.060403,ACSnano.17.20939-20948}.
A fundamental understanding of the basic physics governing filled-gel rheology is, therefore, crucial.


The parameter space of gel composites is huge, and industrial formulations vary from product to product\,\cite{SoftMatter.8.2348-2365,IndEngChemRes.61.2100-2109}.
In this work, we focus on a minimal model system: a strongly aggregating colloidal gel with embedded large, non-sticky fillers.
In experiments, we observe an anomalous inconsistency between the $\phi_{\mathrm{f}}$ dependence of stiffness, represented by elastic (or storage) modulus $G'$, and strength, represented by yield stress $\sigma_{\mathrm{y}}$.
Simulations corroborate this finding, allowing us to unravel the different roles of filling in the above two cases. 
Incorporating a lengthscale argument, our results lead to a mechanistic description that not only has important practical implications (e.g., in battery processing\,\cite{larsen2023controlling} and geophysical flow\,\cite{pradeep2023origins}), but also suggests a conceptual pathway to tunable soft composites.


\section*{Results and Discussions}

\subsection*{Rheological inconsistency}

\begin{figure}[tb]
\centering
\includegraphics[width=\linewidth]{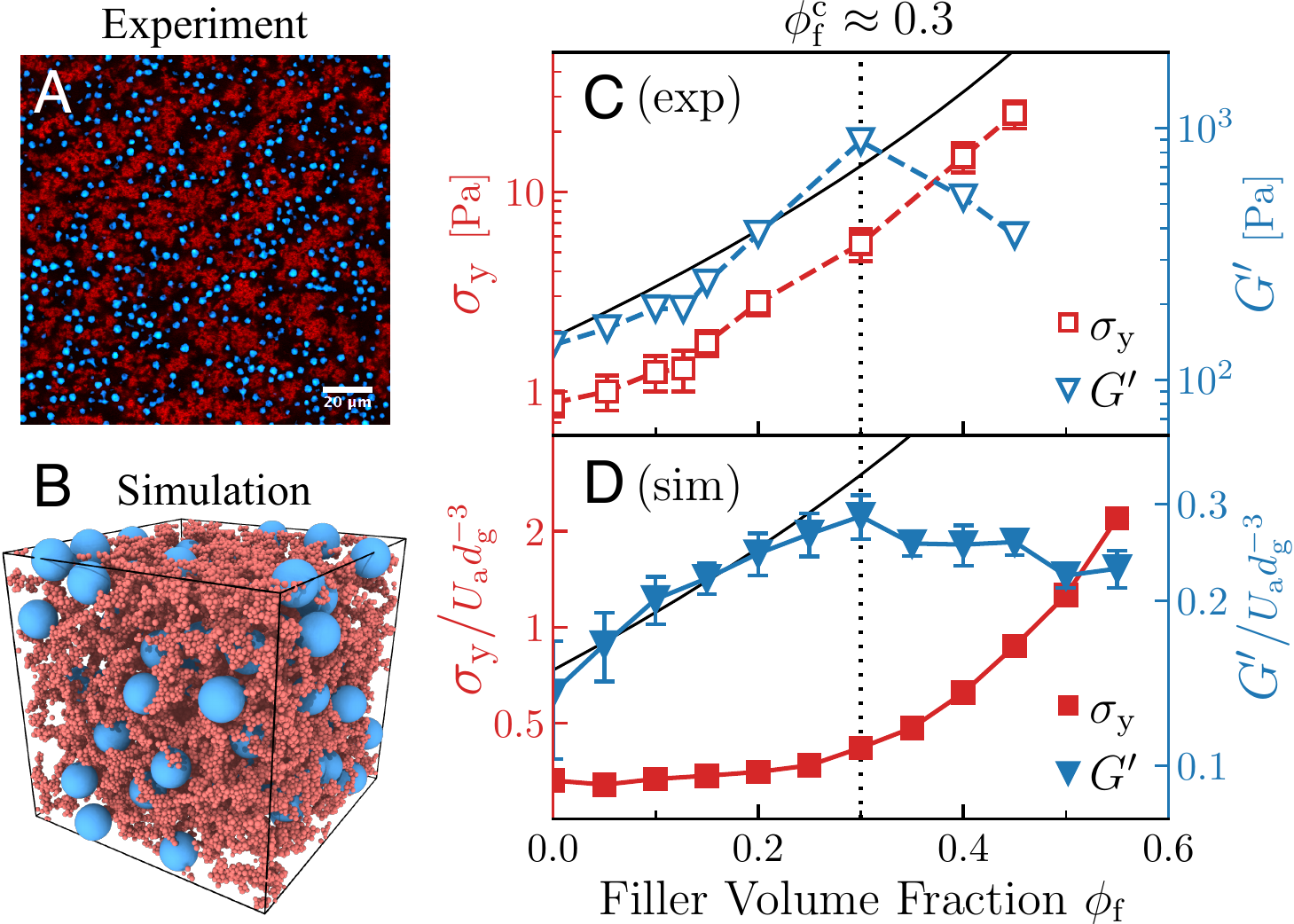}
\caption{
Granular-filled colloidal gels and inconsistency in their rheology.
(\subfig{A})~Colloidal gels with particle fillers in experiments (confocal slice). Red: gel colloids with $\phi_{\mathrm{g}} \approx 0.1$; blue: solid fillers with $\phi_{\mathrm{f}} \approx 0.2$.
(\subfig{B})~3D rendering of a simulated state in similar conditions.
(\subfig{C})~Experimental data of yield stress $\sigma_{\mathrm{y}}$ and elastic modulus $G^\prime$ are plotted as functions of filler volume fraction $\phi_{\mathrm{f}}$. 
The solid black line refers to the $\phi_{\mathrm{eff}}$-predicted elastic moduli (the power-law fitting on unfilled gel data, \sfigref{1}), giving 
$G^\prime =  2.27 \times 10^6 \phi_{\mathrm{eff}}^{4.22}
\,\si{\pascal}$.
(\subfig{D})~Simulation data of $\sigma_{\mathrm{y}}$ and $G^\prime$. 
The solid black line shows 
$G^\prime/U_{\mathrm{a}}d_{\mathrm{g}}^{-3} = 29.8\, \phi_{\mathrm{eff}}^{2.31}$ 
(The fitting on unfilled gel data, \sfigref{2}).
}
\label{fig1}
\end{figure}

Our experimental model system consists of two batches of silica particles differing in size and hydrophobicity (\matref). 
Dispersed in an aqueous solvent with fillers (silica microspheres) of diameter $d_{\mathrm{f}} \approx \SI{4}{\micro\meter}$, the hydrophobic silica (i.e., gel colloids) of diameter $d_{\mathrm{g}} \approx \SI{482}{\nano\meter}$ strongly attract each other ($U_\mathrm{a} \gg k_\mathrm{B}T$) and form a gel network with the fillers embedded inside, \figrefp{fig1}{A}.
We fix the volume fraction of gel colloids at $\phi_\mathrm{g}=0.1$ and vary the volume fraction of fillers $\phi_\mathrm{f}$.
To ensure an intact gelled state, we always perform a high-rate rejuvenation at $\dot{\gamma}=\SI{1000}{\per\second}$ and subsequent \SI{30}{\minute}-recovery at rest before rheological measurements.
Oscillatory rheology (at $\gamma_{0}=\SI{0.1}{\percent}$ and $\omega = \SI{1}{\radian\per\second}$) and creep tests (with increasing stress) are then carried out to measure the elastic modulus $G^\prime$ and yield stress $\sigma_{\mathrm{y}}$, respectively.


For unfilled colloidal gels, higher volume fraction $\phi_\mathrm{g}$ leads to the increases in both $G^\prime$ and $\sigma_{\mathrm{y}}$ (\sfigref{1}).
According to the effective gel volume fraction $\phi_{\mathrm{eff}}$ defined in \eqref{phi_eff}, therefore, solid filling is expected to reinforce the entire system.
Indeed, at fixed $\phi_\mathrm{g}=0.1$, the yield stress $\sigma_{\mathrm{y}}$ (red) increases monotonically with the filler volume fraction $\phi_{\mathrm{f}}$, \figrefp{fig1}{C}.
However, the elastic modulus $G^\prime$ (blue) exhibits a peak at $\phi_{\mathrm{f}}^{\mathrm{c}} \approx 0.3$ and then decreases, i.e., softening.
Such a drop in modulus with dense filling conflicts with the
expectation from increasing $\phi_{\mathrm{eff}}$ (solid black line), and deviates markedly from the trend of increasing yield strength.


This rheological inconsistency between $\sigma_{\mathrm{y}}$ and $G^\prime$ is also observed in simulations.
We simulate binary collections of attractive gel colloids with diameter $d_{\mathrm{g}}$ and non-sticky fillers with diameter $d_{\mathrm{f}}=8d_{\mathrm{g}}$, following the experimental conditions, $\phi_{\mathrm{g}}=0.1$, \figrefp{fig1}{B}.
Both volume exclusion and attraction are modeled using linear Hookean springs with stiffness $k$.
Attraction is set to be short-ranged ($\zeta =0.01d_{\mathrm{g}}$) and strong ($U_{\mathrm{a}} \equiv k\zeta^2/2 = 20k_{\mathrm{B}}T$)\footnote{Though still weaker than the hydrophobic attraction in experiments ($\gtrsim 10^2k_\mathrm{B}T$), we confirm that such a value causes irreversible bonding and is sufficient to produce a fractal gel network, similar to that resulting from stronger attraction.}.
The consequent stress $U_{\mathrm{a}}d_{\mathrm{g}}^{-3}$ is used to scale all the simulation quantities in units of \textit{energy/length$^{3}$} unless otherwise stated.


Gelation begins from a random configuration and proceeds over $10^3 \tau_{\mathrm{B}}$ under a Langevin thermostat until reaching a stable gel, where $\tau_{\mathrm{B}} \equiv \pi\eta d_{\mathrm{g}}^3/2k_{\mathrm{B}}T$ is the Brownian time scale of gel colloids with solvent viscosity $\eta$.
We then relax the gel by gradual quenching to $k_{\mathrm{B}}T=0$.
Our primary concern in this study is the effect of fillers on the network morphology and the resulting mechanics.
As we consider sufficiently strong colloidal bonding, dynamical effects (e.g., caging effect\,\cite{JChemPhys.128.084509} and glass transition\,\cite{Langmuir.28.1866-1878}) contribute little to the rheology.
Therefore, our simulations measure the rheology at athermal conditions following a full quench.
The relevant flow rates ($\gamma_{0} \omega $ in oscillatory rheology and $\dot{\gamma}$ in creep test) remain sufficiently small to eliminate inertial effects (\matref).
Similar to our experimental results in \figrefp{fig1}{C}, we find that the yield stress $\sigma_{\mathrm{y}}$ monotonically increases with $\phi_{\mathrm{f}}$ while the elastic modulus $G^\prime$ first increases and then decreases, \figrefp{fig1}{D}.
The critical filler content $\phi_{\mathrm{f}}^{\mathrm{c}} \approx 0.3$, where $G'$ peaks, is consistent with the experimental result.


\subsection*{Non-monotonic Elastic Modulus}

\begin{figure}
\centering
\includegraphics[width=\linewidth]{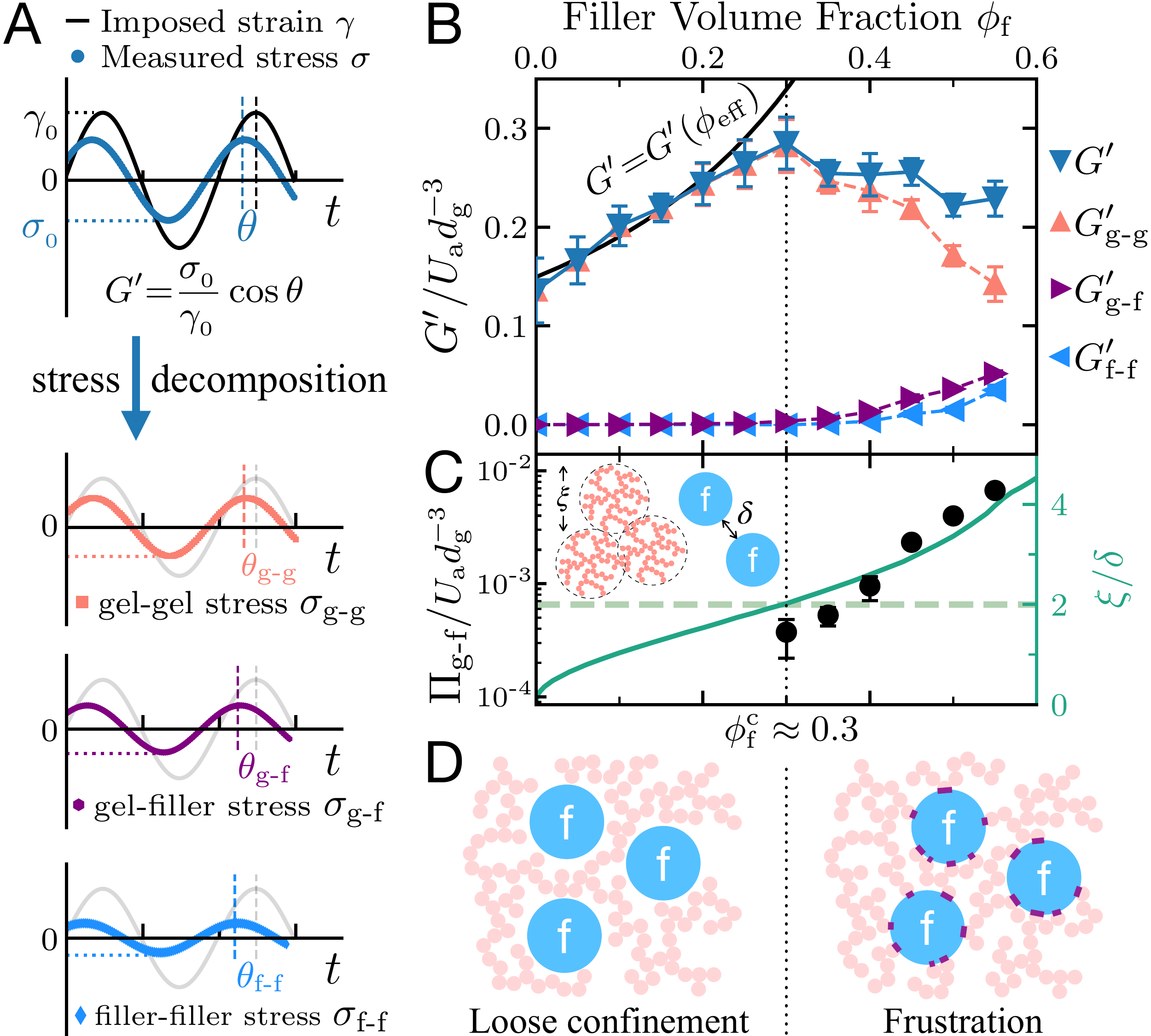}
\caption{
Decomposition of elastic modulus $G^\prime$.
(\subfig{A})~Sketch of stress decomposition in oscillatory simulations.
(\subfig{B})~Decomposed elastic moduli $G^\prime$ as functions of filler volume fraction $\phi_{\mathrm{f}}$. 
The geometric means and standard deviations are plotted here and later.
Solid line: $G^\prime/U_{\mathrm{a}}d_{\mathrm{g}}^{-3} = 29.8 \, \phi_{\mathrm{eff}}^{2.31}$ with $\phi_{\mathrm{eff}}$ defined in \eqref{phi_eff}. 
(\subfig{C})~Inter-phase pressure $\Pi_{\text{g-f}}$ (\eqref{stress}) and lengthscale ratio $\xi/\delta$ as functions of $\phi_{\mathrm{f}}$. 
$\Pi_{\text{g-f}}$ is nonzero for $\phi_{\mathrm{f}} \geq \phi_{\mathrm{f}}^{\mathrm{c}}$.
The gel length $\xi/d_{\mathrm{g}} = 1.44 \, \phi_{\mathrm{eff}}^{-0.88}$ is extrapolated from the fitting of unfilled gel data (\sfigref{3}), while the filler spacing $\delta$ is measured from individual simulations (\sfigref{5}).
The dashed green line refers to $\xi/\delta=2$, and the inset illustrates the physical meanings of $\xi$ and $\delta$. 
(\subfig{D})~Schematic gel matrices at low $\phi_{\mathrm{f}} < \phi_{\mathrm{f}}^{\mathrm{c}}$ (left: loosely confined) and high $\phi_{\mathrm{f}} \geq \phi_{\mathrm{f}}^{\mathrm{c}}$ (right: frustrated). Inter-phase contacts are highlighted in purple.
}
\label{fig2}
\end{figure}

The slow flow in rheological simulations enables us to ignore the kinetic contribution ($\propto mv^2$) and bulk viscous stress  ($\propto \eta\dot{\gamma}$) in the shear stress $\sigma$.
To gain mechanistic insight, we decompose $\sigma$ into contributions from each of three interacting pairs:
\begin{equation}
\sigma = \sigma_{\text{g-g}} + \sigma_{\text{g-f}} + \sigma_{\text{f-f}},\label{decompose}
\end{equation}
then obtain the corresponding moduli $G^\prime_{\text{g-g}}$, $G^\prime_{\text{g-f}}$, and $G^\prime_{\text{f-f}}$ by Fourier transformation of the respective signals, \figrefp{fig2}{A}.
The modulus decomposition shows that, in the dilute regime ($\phi_{\mathrm{f}} \lesssim \phi_{\mathrm{f}}^{\mathrm{c}}$), the elastic response is dominated by that of the gel matrix $G^\prime_{\text{g-g}}$, \figrefp{fig2}{B}.
Here, fillers are mostly isolated, or \emph{loosely confined}, within the gel matrix as shown in \figrefp{fig2}{D} (left).
The occasional gel-filler and filler-filler contacts are unsustainable and thus respond viscously, leading to vanishing $G^\prime_{\text{g-f}}$ and $G^\prime_{\text{f-f}}$ in \figrefp{fig2}{B}.


Extending the power law of unfilled gels (\sfigref{2}) to filled gels, we find that the elastic modulus predicted using $\phi_{\mathrm{eff}}$ in place of $\phi_\mathrm{g}$ (solid black line in \figrefp{fig2}{B}) appears to capture well the measured $G^\prime$ below $\phi_{\mathrm{f}}^{\mathrm{c}}$.
That is, consistent with accelerated gelation\,\cite{NatCommun.14.2773}, fillers concentrate the gel matrix in a manner effectively described by $\phi_{\mathrm{eff}}$.
While such $G^\prime$--$\phi_{\mathrm{eff}}$ scaling is expected locally, this is not trivial on a global level since loosely-confined fillers generate cavities in the gel network.
Presumably, both the stress and strain are associated with their local counterparts by a factor of $\sim (1-\phi_{\mathrm{f}})^{-1}$\,\cite{JFluidMech.852.P1} so that they cancel out in $G^\prime$.


As the filling content increases ($\phi_{\mathrm{f}} \gtrsim \phi_{\mathrm{f}}^{\mathrm{c}}$), fillers start to \emph{frustrate} and thereby soften the gel skeleton, \figrefp{fig2}{D} (right).
This is evidenced by the weakened gel stiffness $G^\prime_{\text{g-g}}$, which decays faster than the overall elastic modulus $G^\prime$.
While the gel-matrix contribution still dominates, the filler-induced moduli $G^\prime_{\text{g-f}}$ and $G^\prime_{\text{f-f}}$ emerge above $\phi_{\mathrm{f}}^{\mathrm{c}}$ and increase with filling, \figrefp{fig2}{B}.
This implies that the fillers are persistently trapped within the gel skeleton via mechanical contacts, which enable them to participate directly in structural percolation and stress transmission.


In contrast to dilutely filled gels ($\phi_{\mathrm{f}} < 0.3$), fillers participating in the load-bearing network serve as junctions and thus induce stress concentration due to the size disparity.
The stress transmission is, therefore, heterogeneous\,\cite{JCIS.598.56-68}.
In addition, the gel-filler and filler-filler pairs cannot support tensile loads, unlike the gel-gel matrix, where both tension and compression generate elastic forces.
Together, these factors result in less efficient stress transmission and, therefore, a reduction in elastic modulus $G^\prime$.

Moreover, the mechanical frustration from gel-filler contacts above $\phi_{\mathrm{f}}^{\mathrm{c}}$ plays a role.
We characterize this frustration using the inter-phase particle pressure
\begin{equation}
\Pi_{\text{g-f}} \equiv 
-\frac{1}{3}
\tr (
\mathbf{\Sigma}_{\text{g-f}}
),
\label{stress}
\end{equation}
where $\mathbf{\Sigma}_{\text{g-f}}$ denotes the gel-filler stress tensor measured at rest.
At lower filler fractions $\phi_{\mathrm{f}} \lesssim \phi_{\mathrm{f}}^{\mathrm{c}}$, the vanishing inter-phase pressure $\Pi_{\text{g-f}} = 0 $ (\figrefp{fig2}{C}) is consistent with the absence of $G^\prime_{\text{g-f}}$ in \figrefp{fig2}{B}, again confirming loosely confined fillers.
Note that the geometric means of five independent simulations are plotted; if the data contains $0$, it is $0$.


At high filling $\phi_{\mathrm{f}} \gtrsim \phi_{\mathrm{f}}^{\mathrm{c}}$ and with sufficient relaxation prior to measurement, $\Pi_{\text{g-f}}$, or equivalently the frustration on the gel, cannot be fully relaxed due to mechanical trapping.
Instead, it becomes non-zero beyond $\phi_{\mathrm{f}}^{\mathrm{c}}\approx0.3$ and increase with $\phi_{\mathrm{f}}$, \figrefp{fig2}{C}.
This coincides with the observed finite elastic responses from solid fillers ($G^\prime_{\text{g-f}}$ and $G^\prime_{\text{f-f}}$). 
Such filler-induced prestress also indicates another possible origin of softening in $G^\prime$\,\cite{PhysRevRes.4.043181}.


A simple lengthscale argument seems to capture the deviation in $G^\prime$ from the $\phi_{\mathrm{eff}}$ prediction reported here. 
For colloidal gels, a characteristic lengthscale $\xi$ can be estimated from the static structure factor $S\left(q\right)$\,\cite{AdvColloidInterfaceSci.73.71-126}. 
As $\xi$ decreases with $\phi_\mathrm{g}$ according to a power-law as $\xi/d_{\mathrm{g}} = 1.44 \phi_\mathrm{g}^{-0.88}$ (\sfigref{3}), we assume such a scaling holds in filled gels by replacing $\phi_\mathrm{g}$ with $\phi_{\mathrm{eff}}$.
Another relevant lengthscale is the average spacing between fillers $\delta$, which has been verified important in various gel composites\,\cite{SoftMatter.18.2842-2850,Macromolecules.51.5375-5391}.
Through Voronoi analysis, we show how $\delta$ decreases with $\phi_{\mathrm{f}}$ (\sfigref{5}).
When their ratio $\xi/\delta$ is small, aggregated clusters can be naturally accommodated between fillers without being frustrated, forming a colloidal gel network well characterized by $\phi_{\mathrm{eff}}$.
As the ratio increases, the reduction in free space leads to inevitable frustration on the gel matrix.
A critical value $\xi/\delta=2$ (dashed green line), which captures the deviation of gelation time in Ref.\,\cite{NatCommun.14.2773}, also coincides with the transition of $G^\prime$ away from the unfilled prediction, as well as the emergent inter-phase particle pressure $\Pi_{\text{g-f}}$, at $\phi_{\mathrm{f}}^{\mathrm{c}} \approx 0.3$, \figsrefp{fig2}{B} and \textit{C}.
Importantly, this lengthscale argument holds for various filler sizes $d_{\mathrm{f}}$ (\sfigref{6}), suggesting a general physics of lengthscale interplay.


\subsection*{Yielding Resistance}

\begin{figure*}
\centering
\includegraphics[width=0.9\textwidth]{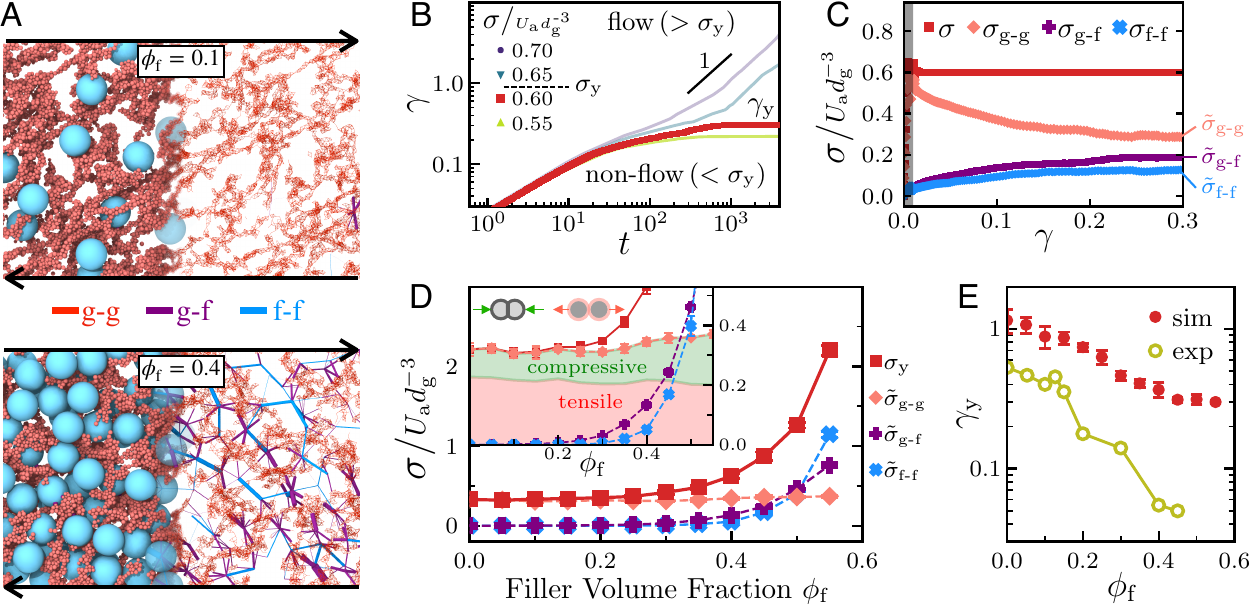}
\caption{
Yielding resistance.
(\subfig{A})~Interacting pairs of a dilutely-filled gel (upper, $\phi_{\mathrm{f}} = 0.1$) and a densely-filled gel (lower, $\phi_{\mathrm{f}} = 0.4$) prior to yielding ($\gamma\to\gamma_{\mathrm{y}}^{-}$). 
Pair types are represented by different colors of bonds, whose width is proportional to the force magnitude. 
Slices with a thickness of $10d_{\mathrm{g}}$ in the flow-gradient plane are shown. 
See full movies in \smovsref{1}{2}.
(\subfig{B})~Creep test results of a filled gel at $\phi_{\mathrm{f}}=0.4$. 
Under a series of imposed stress $\sigma$, the strain $\gamma$ evolves over time, either reaching a finite value (non-flow, $\sigma<\sigma_{\mathrm{y}}$) or linearly growing (flow, $\sigma>\sigma_{\mathrm{y}}$).
(\subfig{C})
For the data just below the yield stress ($\sigma/U_{\mathrm{a}}d_{\mathrm{g}}^{-3}=0.60$, red), the stress $\sigma$ is further decomposed by pair species (\eqref{decompose}), and their evolutions over strain $\gamma$.
(\subfig{D})~Yield stress $\sigma_{\mathrm{y}}$ and contributions from each interacting pair type ($\tilde{\sigma}_{\text{g-g}}$, $\tilde{\sigma}_{\text{g-f}}$, and $\tilde{\sigma}_{\text{f-f}}$) as functions of filler volume fraction $\phi_{\mathrm{f}}$. Inset shows further decomposition of $\tilde{\sigma}_{\text{g-g}}$ into compressive and tensile loads.
(\subfig{E})~Yield strains $\gamma_{\mathrm{y}}$ (determined by creep tests) from simulations (sim) and experiments (exp) as functions of filler volume fraction $\phi_{\mathrm{f}}$.
}
\label{fig3}
\end{figure*}

Although mechanical frustration caused by dense filling reduces the stiffness, it does not affect the yielding resistance; the yield stress $\sigma_{\mathrm{y}}$ monotonically increases with $\phi_{\mathrm{f}}$, \figsrefp{fig1}{C} and \textit{D}.
Creep tests illustrate the microscopic origin. 
Through stress-controlled simulations (\matref), we examine the strain evolutions $\gamma\left(t\right)$ under a series of stresses $\sigma$ (\figrefp{fig3}{B}), by which the yield stress $\sigma_{\mathrm{y}}$ and yield strain $\gamma_{\mathrm{y}}$ can be determined.
For the simulation just below yielding ($\sigma \to \sigma_{\mathrm{y}}^{-}$), stress decomposition (\eqref{decompose}) is applied to show the evolution of each stress component, \figrefp{fig3}{C}.
We regard the stress value in the steady state ($\gamma \to \gamma_{\mathrm{y}}$) as the contribution of each interaction type to the overall yield strength: $\tilde{\sigma}_{\text{g-g}}$, $\tilde{\sigma}_{\text{g-f}}$, and $\tilde{\sigma}_{\text{f-f}}$.


For filled gels with low $\phi_{\mathrm{f}}$ under creeping, the stress is mostly undertaken by the gel matrix (\figrefp{fig3}{A}, upper), i.e., $\tilde{\sigma}_{\text{g-g}} \approx \sigma_{\mathrm{y}}$ as shown in \figrefp{fig3}{D}.
This supports our argument that fillers are loosely confined and, therefore, their stress can be relaxed under load prior to yielding.
As $\phi_{\mathrm{f}}$ increases, the other two components ($\tilde{\sigma}_{\text{g-f}}$ and $\tilde{\sigma}_{\text{f-f}}$) grow and finally exceed $\tilde{\sigma}_{\text{g-g}}$, \figrefp{fig3}{D}. 
In a highly filled gel, $\phi_{\mathrm{f}}=0.4$, the fillers develop into percolating force chains along the compressive direction under shear (\figrefp{fig3}{A}, lower) and resist yielding as a result.
Further decomposing the gel response $\tilde{\sigma}_{\text{g-g}}$ into tensile and compressive components, we find that the ratio of compressive loads increases with $\phi_{\mathrm{f}}$ (\textit{Inset} of \figrefp{fig3}{D}).


Surprisingly, in contrast to the filling-dependent $G'$ shown in \figsref{fig1} and \ref{fig2}, the yield stress contribution $\tilde{\sigma}_{\text{g-g}}$ from the gel matrix remains almost independent of $\phi_{\mathrm{f}}$, \figrefp{fig3}{D}.
In other words, solid filling either enhances (loose confinement) or weakens (mechanical frustration) the stiffness $G^\prime_{\text{g-g}}$ of gel matrix depending on $\phi_{\mathrm{f}}$, but seemingly plays little role in the yielding resistance $\tilde{\sigma}_{\text{g-g}}$.
This distinction is associated with the physical difference between the two quantities.
The elastic modulus $G^\prime$ represents the stiffness of a material (i.e., how difficult to deform), and is typically measured at a small strain amplitude ($\gamma_0 = \SI{0.1}{\percent}$ in this work).
In contrast, yield stress $\sigma_{\mathrm{y}}$, or strength, refers to the maximum load that a material can bear prior to failure, which usually occurs far beyond the linear regime ($\gamma_{\mathrm{y}} \gtrsim \SI{10}{\percent}$ in this work).
We conclude, therefore, that the monotonic strengthening in $\sigma_{\mathrm{y}}$ stems from the additional mechanical support, mainly compressive resistance, offered by solid fillers.


\subsection*{Filler-Removal and Tunability}

\begin{SCfigure*}[\sidecaptionrelwidth][t!]
\centering
\includegraphics[width=13.9cm]{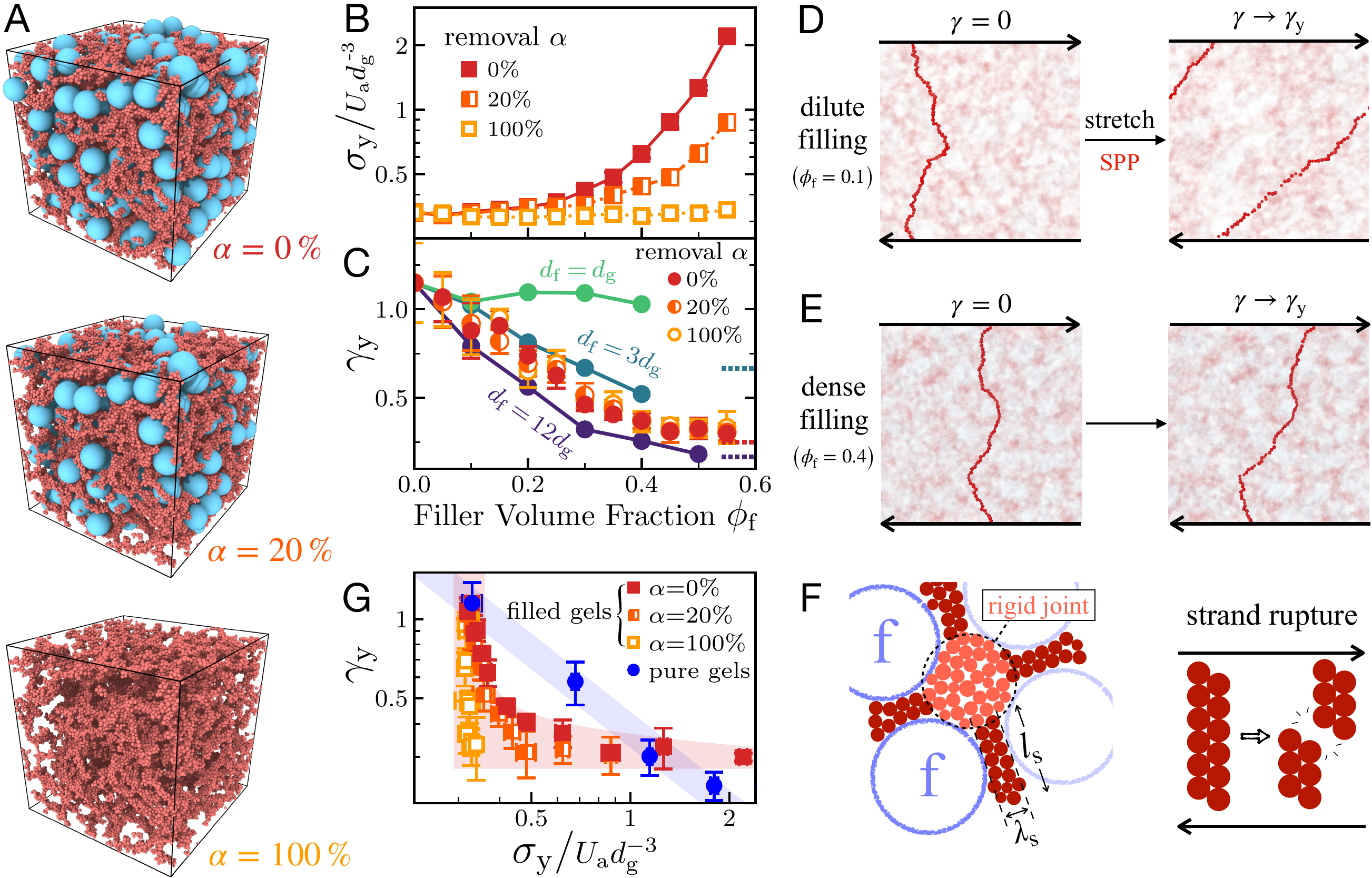}
\caption{
Filler-removal protocol.
(\subfig{A})~Filler-removal sketch with varying removal fraction $\alpha$.
(\subfig{B})~Yield stress $\sigma_{\mathrm{y}}$ as a function of filler volume fraction $\phi_{\mathrm{f}}$ and removal fraction $\alpha$.
(\subfig{C})~Yield strain $\gamma_{\mathrm{y}}$ as a function of filler volume fraction $\phi_{\mathrm{f}}$, removal fraction $\alpha$, and filler size $d_{\mathrm{f}}$.
Dashed lines on the right-hand side refer to the predicted $\gamma_{\mathrm{y}} \sim \lambda_{\mathrm{s}}/l_{\mathrm{s}}$, see (\subfig{F}).
(\subfig{D} and \subfig{E})~Stretched SPP (highlighted in red) under creeping of dilutely-filled gel (\subfig{D}, $\phi_{\mathrm{f}}=0.1$) and densely-filled gel (\subfig{E}, $\phi_{\mathrm{f}}=0.4$). 
See \smovsref{3}{4} for full structural evolutions. 
(\subfig{F})~Schematic strand rupture in densely-filled gels upon yielding.
(\subfig{G})~$\sigma_{\mathrm{y}}$--$\gamma_{\mathrm{y}}$ diagram of colloidal gels (blue) and filled gels upon filler-removal at different $\alpha$.
}
\label{fig4}
\end{SCfigure*}

The transition in yielding mechanism also manifests in the yield strain $\gamma_{\mathrm{y}}$.
Though quantitatively different%
\footnote{The quantitative difference may result from the difference in interaction range and contact mechanics.}%
, both experiments and simulations confirm that solid filling leads to a reduction in $\gamma_{\mathrm{y}}$, \figrefp{fig3}{G}.
This embrittlement indicates more possibilities in tunable gel rheology.
To further extend such tunability, as well as to better illustrate the role of solid filling, here we develop a \emph{filler-removal protocol}.
This protocol is inspired by the template-removal technique\,\cite{ChemCommun.1186-1187} in mesoporous materials.
Specifically, we remove fillers following gelation and perform relaxation and rheological measurements afterward.
The proportion $\alpha$ of fillers to remove (randomly selected) then becomes a tuning parameter, \figrefp{fig4}{A}.


While the yield strength monotonically increases with $\phi_{\mathrm{f}}$, removing all fillers ($\alpha=\SI{100}{\percent}$) always reduces the yield stress $\sigma_{\mathrm{y}}$ to that of an unfilled gel, \figrefp{fig4}{B}.
This is consistent with the insignificant $\phi_{\mathrm{f}}$ dependence of the stress contribution $\tilde{\sigma}_{\text{g-g}}$ from gel matrix, \figrefp{fig3}{D}.
In addition, a gel with partially removed fillers ($\alpha=\SI{20}{\percent}$) exhibits an intermediate yield stress between those of a full composite ($\alpha=\SI{0}{\percent}$) and a filler-removed matrix ($\alpha=\SI{100}{\percent}$), \figrefp{fig4}{B}.
These results, as well as the lower $G^\prime$ upon filler-removal (\sfigref{8}), further elucidate the role of solid fillers in mechanical reinforcement.

Interestingly, the filler-removal operation appears to have little effect on the yield strain $\gamma_{\mathrm{y}}$, \figrefp{fig4}{C}.
This implies the structural origin of filler-induced embrittlement.
Though increasing the concentration typically reduces the yield strain in colloidal gels\,\cite{PhysRevA.42.4772}, (also see \sfigref{2}), the reduction in $\gamma_{\mathrm{y}}$ may not be simply attributed to the increasing $\phi_{\mathrm{eff}}$.
With the same composition, the yield strain $\gamma_{\mathrm{y}}$ varies as a function of the filler size $d_{\mathrm{f}}$.
Larger fillers ($d_{\mathrm{f}} = 12d_{\mathrm{g}}$) give rise to lower $\gamma_{\mathrm{y}}$; comparable sizes ($d_{\mathrm{f}} = d_{\mathrm{g}}$) barely change $\gamma_{\mathrm{y}}$, \figrefp{fig4}{C}.
These results suggest another size-dependent physics, which is not accounted for $\phi_{\mathrm{eff}}$ in \eqref{phi_eff}.


Dilute gels of strongly aggregating colloids consist of ramified clusters\,\cite{AdvColloidInterfaceSci.73.71-126}, in which thin strands are packed in a fractal manner (\sfigref{3}).
With strong attractions, these strands hardly deform on their own\,\cite{PhysRevLett.120.208005}, but their joints, instead, are relatively flexible to allow cluster rotations.
Hence, the gel network evolves into a highly-stretched state under shear\,\cite{JRheol.58.1089-1116}, and the yielding point correlates with the fracture of the most-stretched strand, \figrefp{fig4}{D} and \smovref{3}.
While network morphology is determinative to gel rheology\,\cite{PNAS.121.e2316394121}, a quantitative analysis based on the shortest percolation path (SPP)\,\cite{JStatPhys.93.603-613} agrees with our measurement on $\gamma_{\mathrm{y}}$ for dilute gels (\sfigref{4}).


Filling at high $\phi_{\mathrm{f}}$ solidifies the joint-strand connections \emph{via} squeezing, \figrefp{fig4}{F}.
If assuming fully-constrained joints at the dense limit, yielding occurs with the rupture, rather than stretch (\figrefp{fig4}{E} and \smovref{4}), of the strands.
The yield strain $\gamma_{\mathrm{y}}$ may be alternatively estimated by the aspect ratio of strands.
While the strand length $l_{\mathrm{s}}$ is comparable to the filler size $d_{\mathrm{f}}$ as sketched in \figrefp{fig4}{F} (left), the width $\lambda_{\mathrm{s}} \approx 2d_{\mathrm{g}}$ can be approached by considering a strand of tetrahedrons (the minimal rigid element\,\cite{NatPhys.19.1171-1177}).
Thus, we can estimate the yield strain as $\gamma_{\mathrm{y}} \sim \lambda_{\mathrm{s}}/l_{\mathrm{s}}$, which is $d_{\mathrm{f}}$ dependent.
For large $d_{\mathrm{f}}$, the estimate is in quantitative agreement with the simulations, \figrefp{fig4}{C} (dashed lines). 
Small fillers, nevertheless, behave more like a continuous fluid phase without modifying the final gel structure (\sfigref{7}) and thereby have less impact on $\gamma_{\mathrm{y}}$. 
While the two yielding scenarios we proposed (\figsrefp{fig4}{D--F}) exhibit semi-quantitative agreement with simulational measurements, experimental verifications are, unfortunately, unpractical due to the limited view and time resolution in microscopy.


The disparate dependence of yield stress and strain on filler-removal allows for additional tunability.
For colloidal gels, changing volume fraction merely enables limited exploration on the $\sigma_{\mathrm{y}}$--$\gamma_{\mathrm{y}}$ diagram, \figrefp{fig4}{G} (blue).
With the filler-removal protocol applied, the yield strength $\sigma_{\mathrm{y}}$ varies with both the filler volume fraction $\phi_{\mathrm{f}}$ and the removal fraction $\alpha$, while the yield strain $\gamma_{\mathrm{y}}$ is a function of $\phi_{\mathrm{f}}$ only.
This suggests a novel way to tune gel rheology.
By replotting the data of \figsrefp{fig4}{B} and \textit{C} ($d_{\mathrm{f}}=8d_{\mathrm{g}}$) in the same diagram, we find that filling techniques enable continuous exploration to a more brittle state (low $\sigma_{\mathrm{y}}$ and low $\gamma_{\mathrm{y}}$), \figrefp{fig4}{G} (red), which is not naturally accessible by unfilled gels.
That is, the filler-removal protocol enables individual control over strength and brittleness.


There are several candidate experimental protocols in which this computational proof-of-concept could be realised.
One is to adapt current template-removal techniques, which utilize microwave\,\cite{ChemCommun.1186-1187}, ultraviolet light\,\cite{ChemCommun.1998.2203-2204}, chemical reflux\,\cite{EnergyFuels.36.2424-2446}, etc.
Using fillers of thermo-sensitive materials (e.g., pNIPAAm\,\cite{Biomater.25.3793-3805} and gelatin\,\cite{IntJBioprint.8.599}), filler-removal protocol may also be achieved via temperature control.
Alternatively, bubbles- or droplets-embedded gel composites \cite{TFST.86.85-94,FoodHydrocoll.119.106875,SoftMatter.19.2771,SoftMatter.18.2092} exhibit potential of practical filler removal as well.
While these routes highly enrich the control diagram, tunability usung other quantities, such as gel volume fraction $\phi_{\mathrm{g}}$\,\cite{PhysRevA.42.4772}, filler size $d_{\mathrm{f}}$ (\sfigref{6}), and interparticle friction\,\cite{NatCommun.14.5309}, presents a vast parameter space of rheological control by formulation
that can be made rational by the mechanistic insight provided here.


\section*{Conclusions}

Altogether we probe the impact of solid filling on colloidal gel rheology, focusing on the case of large, non-sticky fillers embedded in a matrix of a dilute, yet strong ($U_{\mathrm{a}} \gg k_{\mathrm{B}}T$), colloidal gel.
With more fillers added to the gel, the inconsistency between yield strength ($\sim \sigma_{\mathrm{y}}$) and stiffness ($\sim G^\prime$) indicates different roles of solid filling. 
Fillers can additionally support the gel matrix to resist further yielding under sustained load and large strain,
whereas beyond $\phi_{\mathrm{f}}^{\mathrm{c}} \approx 0.3$ the structural frustration on the gel backbone dominates and leads to softening under small amplitude oscillatory shear.
This structural impact also leads to a reduction in yield strain $\gamma_{\mathrm{y}}$. 
Exploiting the novel filler-removal protocol described here, one may individually control the strength and brittleness to achieve a new practical route to tunability in colloidal gel rheology.


\matmethods{\label{matmethods}

\subsection*{Experiments}

To fabricate the experimental system, we mixed hydrophobic silica colloids of size $d_{\mathrm{g}} \approx \SI{482}{\nano\meter}$ (from light scattering) and large silica microspheres of diameter $d_{\mathrm{f}} \approx \SI{4}{\micro\meter}$ (Angstromspheres, Blue Helix Ltd.) in an aqueous solvent.
The small colloids are fluorescently labeled by rhodamine B to enable confocal imaging and functionalized with Tri-methylsilyl group to render surface hydrophobicity.
The solvent, composed of water, ethanol, and glycerol with a mass ratio of 1:1:9, is nearly refractive-index-matched and dyed with \SI{0.1}{\milli\mole} fluorescein sodium salt.
On their own, the hydrophobic colloids are strongly attractive and aggregate into a tenuous, space-spanning network, i.e., the colloidal gel.
The large silica, on the other hand, has charged surfaces and thereby repel each other within a short range (Debye length $\approx$ \SI{10}{\nano\meter}).
With confocal microscopy, we do not observe significant aggregation between gel colloids and large silica spheres.


To ensure that rheology is properly measured, we rejuvenate the sample with the high shear rate $\dot{\gamma} = \SI{1000}{\per\second}$ for \SI{1}{\minute} and wait at rest for another \SI{30}{\minute}.
Before each measurement, the rejuvenation--waiting protocol is always applied to prevent the flow-induced phase separation as reported in Ref.\,\cite{PhysRevLett.128.248002}.
Subsequent rheology measurements consist of oscillatory rheology and creep tests. 
In particular, we perform oscillatory rheology in the linear regime with $\gamma_0 = \SI{0.1}{\percent}$ and $\omega = \SI{1}{\radian\per\second}$, and creep tests with a constant imposed stress $\sigma$ increasing from \SI{0.1}{\pascal} to $\sim \sigma_{\mathrm{y}}$ in a stepwise manner.


\subsection*{Simulations}

We perform corresponding simulations on LAMMPS\,\cite{ComputPhysCommun.271.108171}. 
To best mimic our experiments, the simulation system is composed of two species of spherical particles that differ in size and interaction. 
The first species (gel colloids) consists of attractive particles with an average diameter $d_{\mathrm{g}}$ (bidispersed with a size ratio 1:1.3 to prevent crystallization), while the second species (solid fillers) consists of stiff spheres of diameter $d_{\mathrm{f}}=8d_{\mathrm{g}}$.
The particle interactions are all modeled using a linear Hookean spring (extended if attractive) with stiffness $k$.
While the fillers stick neither to themselves nor to the gel (i.e., volume exclusion only), the gel colloids attract each other with a short-ranged ($\zeta=0.01d_{\mathrm{g}}$), strong ($U_{\mathrm{a}}=20k_{\mathrm{B}}T$) attraction.
Such attractive strength is sufficient to induce percolated colloidal gel at a low concentration $\phi_{\mathrm{g}}=0.1$, while higher $U_{\mathrm{a}}$ does not lead to a significant difference.
Unless otherwise stated, all the stress-relevant quantities in this work are scaled by the stress from attraction $U_{\mathrm{a}}d_{\mathrm{g}}^{-3}$ (where $U_{\mathrm{a}} \equiv k\zeta^2/2$).


All simulations contain at least $N_{\mathrm{g}}=\num{40000}$ gel colloids and varying amounts of fillers ($N_{\mathrm{f}} > 50$) in a cubic box with Lees--Edwards periodic boundaries. 
Under the Langevin thermostat, gelation starts from a random configuration and evolves for $\num{1000}\tau_{\mathrm{B}}$ until a stable gelled state is reached.
We then quench the system to $k_{\mathrm{B}}T=0$ at a slow rate for sufficient relaxation, after which rheology measurements are carried out under athermal conditions.
Given the particle mass $m$ and fluid viscosity $\eta$, the relevant timescales are always well separated in the following order:
\begin{equation}
\mathrm{d}t \ll t_{\mathrm{c}} \lesssim t_{\mathrm{d}} \ll \tau_{\mathrm{B}} \ll 1/{\dot{\gamma}(\text{or }1/\omega)},
\end{equation}
where $\mathrm{d}t$ refers to the simulation time step, $t_{\mathrm{c}} \equiv \sqrt{m/k}$ to the collision time, $t_{\mathrm{d}} \equiv m/(3\pi\eta d)$ to the damping time, $\tau_{\mathrm{B}} \equiv \pi\eta d^3 /(2k_{\mathrm{B}}T)$ to the Brownian time and $1/\dot{\gamma}$ (or $1/\omega$) to the rheological timescale.
In this way, our simulations are ensured to be stable and overdamped.
Throughout this work, each simulation runs for five times, with the presented data scatters and error bars representing the average (or geometric mean) and standard deviation, respectively. 
Visualization and part of data analysis use OVITO\,\cite{ModelSimulMaterSciEng.18.015012}.


\subsubsection*{Rheology}

In our simulations, oscillatory rheology is carried out with a small amplitude ($\gamma_0 = \SI{0.1}{\percent}$, same as experiments) and a low frequency ($\omega t_{\mathrm{d}}\ll1$) to eliminate inertial effects.
Oscillation proceeds for five full cycles, with the first two cycles for equilibrating and the last three cycles for averaging measurement.
With the time series of shear stress $\sigma(t)$ and sinusoidal shear strain $\gamma = \gamma_{0} \sin(\omega t)$, we use Fourier Transform to extract the stress amplitude $\sigma_{0}$ and phase difference $\theta$.
The elastic modulus can then be computed based on $G^\prime \equiv (\sigma_0 / \gamma_{0} )\cos{\theta}$.
As illustrated in \figrefp{fig2}{A}, such treatment also applies to the decomposed stresses, giving $G^\prime_{\text{g-g}}$, $G^\prime_{\text{g-f}}$, and $G^\prime_{\text{f-f}}$.
With the kinetic contribution and viscous stress ignored, $G^\prime = G^\prime_{\text{g-g}} + G^\prime_{\text{g-f}}+ G^\prime_{\text{f-f}}$ always holds due to the linearity of Fourier Transform.


Rheological simulations on LAMMPS are intrinsically rate-controlled.
To achieve creep tests, we implement a feedback loop internally to maintain a constant shear stress $\sigma$ by adjusting the shear rate $\dot{\gamma}$ at every single step\,\cite{SoftMatter.15.4150423}.
Given a setpoint $\sigma_{\mathrm{set}}$, we measure the shear stress $\sigma(t)$ and then impose a corresponding shear rate $\dot{\gamma}(t)$ based on the distance to the setpoint, i.e., the error value $e(t) \equiv \sigma(t) - \sigma_{\mathrm{set}}$.
Specifically, a Proportional-Integral (PI) controller is applied as follow:
\begin{equation}
\dot{\gamma}(t) = K_{\mathrm{p}} e(t) + K_{\mathrm{i}} \int_{0}^{t}e(t)\,\mathrm{d}t,
\end{equation}
where $K_{\mathrm{p}}$ and $K_{\mathrm{i}}$ are two control parameters.
With appropriate parameter tuning, such a method can stabilize the stress $\sigma$ around the setpoint (with $\lesssim$ \SI{1}{\percent} error) in 20 simulation steps, \sfigref{2}.
By imposing a series of shear stresses, we examine the evolution of strain $\gamma(t)$ to determine the flow ($\gtrsim \sigma_{\mathrm{y}}$) and non-flow ($\lesssim \sigma_{\mathrm{y}}$) states, such as \figrefp{fig3}{B}.
In this way, we can determine both the yield stress $\sigma_{\mathrm{y}}$ (the average of the highest $\sigma$ to induce flow and the lowest $\sigma$ to induce non-flow state) and the yield strain $\gamma_{\mathrm{y}}$ (stable $\gamma$ plateau under the highest $\sigma$ to induce non-flow state).

}
\showmatmethods{}

\acknow{%
We thank John Royer and Wilson Poon for experimental assistance, George Petekidis and Tiancong Zhao for fruitful discussions.
This work was supported by National Natural Science Foundation of China (12174390 and 12150610463) and Wenzhou Institute, University of Chinese Academy of Sciences (WIUCASQD2020002).
Y.J.~acknowledges funding from China Postdoctoral Science Foundation (2022M723114). 
C.N.~acknowledges support from the Royal Academy of Engineering under the Research Fellowship scheme and from the Leverhulme Trust under Research Project Grant RPG-2022-095.%
}
\showacknow{}

\bibsplit[13]


\begin{thebibliography}{10}

\bibitem{JPhysCondensMatter.33.453022}
CP Royall, MA Faers, SL Fussell, JE Hallett, Real space analysis of colloidal
  gels: triumphs, challenges and future directions.
\newblock {\em\protect\JournalTitle{J. Phys. Condens. Matter}} \textbf{33},
  453002 (2021).

\bibitem{RevModPhys.89.035005}
D Bonn, MM Denn, L Berthier, T Divoux, S Manneville, Yield stress materials in
  soft condensed matter.
\newblock {\em\protect\JournalTitle{Rev. Mod. Phys.}} \textbf{89}, 035005
  (2017).

\bibitem{PhysRevLett.106.248303}
J Sprakel, SB Lindstr\"om, TE Kodger, DA Weitz, Stress enhancement in the
  delayed yielding of colloidal gels.
\newblock {\em\protect\JournalTitle{Phys. Rev. Lett.}} \textbf{106}, 248303
  (2011).

\bibitem{NatPhys.19.1178}
M Bantawa, et~al., The hidden hierarchical nature of soft particulate gels.
\newblock {\em\protect\JournalTitle{Nat. Phys.}} \textbf{19}, 1178--1184
  (2023).

\bibitem{PNAS.118.e2022339118}
B Keshavarz, et~al., Time–connectivity superposition and the gel/glass
  duality of weak colloidal gels.
\newblock {\em\protect\JournalTitle{Proc. Natl. Acad. Sci. U.S.A.}}
  \textbf{118}, e2022339118 (2021).

\bibitem{Nature.505.382-385}
S Rose, et~al., Nanoparticle solutions as adhesives for gels and biological
  tissues.
\newblock {\em\protect\JournalTitle{Nature}} \textbf{505}, 382--385 (2014).

\bibitem{FaradayDiscuss.158.9-35}
J Ubbink, Soft matter approaches to structured foods: from ``cook-and-look'' to
  rational food design?
\newblock {\em\protect\JournalTitle{Faraday Discuss.}} \textbf{158}, 9--35
  (2012).

\bibitem{JohnWiley.9781119220510}
ED Gado, et~al., {\em Colloidal Gelation}.
\newblock (John Wiley \& Sons, Ltd), pp. 279--291 (2016).

\bibitem{PhysRevLett.95.238302}
S Manley, et~al., Glasslike arrest in spinodal decomposition as a route to
  colloidal gelation.
\newblock {\em\protect\JournalTitle{Phys. Rev. Lett.}} \textbf{95}, 238302
  (2005).

\bibitem{NatMater.7.556-561}
C Patrick~Royall, SR Williams, T Ohtsuka, H Tanaka, Direct observation of a
  local structural mechanism for dynamic arrest.
\newblock {\em\protect\JournalTitle{Nat. Mater.}} \textbf{7}, 556--561 (2008).

\bibitem{PNAS.109.16029-16034}
LC Hsiao, RS Newman, SC Glotzer, MJ Solomon, Role of isostaticity and
  load-bearing microstructure in the elasticity of yielded colloidal gels.
\newblock {\em\protect\JournalTitle{Proc. Natl. Acad. Sci. U.S.A.}}
  \textbf{109}, 16029--16034 (2012).

\bibitem{NatPhys.2024}
Y Wang, M Tateno, H Tanaka, Distinct elastic properties and their origins in
  glasses and gels.
\newblock {\em\protect\JournalTitle{Nat. Phys.}} (2024).

\bibitem{Nature.453.499-503}
PJ Lu, et~al., {Gelation of particles with short-range attraction}.
\newblock {\em\protect\JournalTitle{Nature}} \textbf{453}, 499--503 (2008).

\bibitem{SoftMatter.11.4640-4648}
N Koumakis, et~al., Tuning colloidal gels by shear.
\newblock {\em\protect\JournalTitle{Soft Matter}} \textbf{11}, 4640--4648
  (2015).

\bibitem{PhysRevX.10.011028}
T Gibaud, et~al., Rheoacoustic gels: Tuning mechanical and flow properties of
  colloidal gels with ultrasonic vibrations.
\newblock {\em\protect\JournalTitle{Phys. Rev. X}} \textbf{10}, 011028 (2020).

\bibitem{PhysRevLett.131.018301}
M Wei, MY Ben~Zion, O Dauchot, Reconfiguration, interrupted aging, and enhanced
  dynamics of a colloidal gel using photoswitchable active doping.
\newblock {\em\protect\JournalTitle{Phys. Rev. Lett.}} \textbf{131}, 018301
  (2023).

\bibitem{ACSNano.13.560-572}
AK Omar, Y Wu, ZG Wang, JF Brady, Swimming to stability: Structural and
  dynamical control via active doping.
\newblock {\em\protect\JournalTitle{ACS Nano}} \textbf{13}, 560--572 (2019).

\bibitem{MaterTodayProc.45.2536-2539}
M {Muhammed Shameem}, S Sasikanth, R Annamalai, R {Ganapathi Raman}, A brief
  review on polymer nanocomposites and its applications.
\newblock {\em\protect\JournalTitle{Mater. Today: Proc.}} \textbf{45},
  2536--2539 (2021).

\bibitem{JMaterSci.19.473-486}
J Spanoudakis, RJ Young, Crack propagation in a glass particle-filled epoxy
  resin.
\newblock {\em\protect\JournalTitle{J. Mater. Sci.}} \textbf{19}, 473--486
  (1984).

\bibitem{MaterStruct.37.74-81}
H Moosberg-Bustnes, B Lagerblad, E Forssberg, The function of fillers in
  concrete.
\newblock {\em\protect\JournalTitle{Mater. Struct.}} \textbf{37}, 74--81
  (2004).

\bibitem{SoftMatter.10.13-38}
V Ganesan, A Jayaraman, Theory and simulation studies of effective
  interactions{,} phase behavior and morphology in polymer nanocomposites.
\newblock {\em\protect\JournalTitle{Soft Matter}} \textbf{10}, 13--38 (2014).

\bibitem{JRheol.52.287-313}
F Mahaut, X Chateau, P Coussot, G Ovarlez, Yield stress and elastic modulus of
  suspensions of noncolloidal particles in yield stress fluids.
\newblock {\em\protect\JournalTitle{J. Rheol.}} \textbf{52}, 287--313 (2008).

\bibitem{PhysRevLett.128.248002}
Y Jiang, S Makino, JR Royer, WCK Poon, Flow-switched bistability in a colloidal
  gel with non-brownian grains.
\newblock {\em\protect\JournalTitle{Phys. Rev. Lett.}} \textbf{128}, 248002
  (2022).

\bibitem{SoftMatter.19.1342-1347}
Y Li, JR Royer, J Sun, C Ness, Impact of granular inclusions on the phase
  behavior of colloidal gels.
\newblock {\em\protect\JournalTitle{Soft Matter}} \textbf{19}, 1342--1347
  (2023).

\bibitem{NatCommun.14.2773}
Y Jiang, R Seto, Colloidal gelation with non-sticky particles.
\newblock {\em\protect\JournalTitle{Nat. Commun.}} \textbf{14}, 2773 (2023).

\bibitem{SoftMatter.18.2842-2850}
C Ferreiro-C{\'o}rdova, et~al., Stiffening colloidal gels by solid inclusions.
\newblock {\em\protect\JournalTitle{Soft Matter}} \textbf{18}, 2842--2850
  (2022).

\bibitem{Langmuir.34.3021-3029}
D Jia, H Cheng, CC Han, Interplay between caging and bonding in binary
  concentrated colloidal suspensions.
\newblock {\em\protect\JournalTitle{Langmuir}} \textbf{34}, 3021--3029 (2018).

\bibitem{SoftMatter.11.8818-8826}
D Jia, JV Hollingsworth, Z Zhou, H Cheng, CC Han, Coupling of gelation and
  glass transition in a biphasic colloidal mixture---from gel-to-defective
  gel-to-glass.
\newblock {\em\protect\JournalTitle{Soft Matter}} \textbf{11}, 8818--8826
  (2015).

\bibitem{AdvEnergyMater.1.511-516}
M Duduta, et~al., Semi-solid lithium rechargeable flow battery.
\newblock {\em\protect\JournalTitle{Adv. Energy Mater.}} \textbf{1}, 511--516
  (2011).

\bibitem{PhysRevE.77.060403}
A Mohraz, ER Weeks, JA Lewis, Structure and dynamics of biphasic colloidal
  mixtures.
\newblock {\em\protect\JournalTitle{Phys. Rev. E}} \textbf{77}, 060403 (2008).

\bibitem{ACSnano.17.20939-20948}
I Dellatolas, et~al., Local mechanism governs global reinforcement of
  nanofiller-hydrogel composites.
\newblock {\em\protect\JournalTitle{ACS Nano}} \textbf{17}, 20939--20948
  (2023).

\bibitem{SoftMatter.8.2348-2365}
D Das, T Kar, PK Das, Gel-nanocomposites: materials with promising
  applications.
\newblock {\em\protect\JournalTitle{Soft Matter}} \textbf{8}, 2348--2365
  (2012).

\bibitem{IndEngChemRes.61.2100-2109}
JH Park, SH Sung, S Kim, KH Ahn, Significant agglomeration of conductive
  materials and the dispersion state change of the ni-rich nmc-based cathode
  slurry during storage.
\newblock {\em\protect\JournalTitle{Ind. Eng. Chem. Res.}} \textbf{61},
  2100--2109 (2022).

\bibitem{larsen2023controlling}
T Larsen, et~al., Controlling the rheo-electric properties of graphite/carbon
  black suspensions by `flow switching'.
\newblock {\em\protect\JournalTitle{Rheol. Acta}} (2024).

\bibitem{pradeep2023origins}
S Pradeep, PE Arratia, DJ Jerolmack, Origins of complexity in the rheology of
  soft earth suspensions (2023).

\bibitem{JChemPhys.128.084509}
DC Viehman, KS Schweizer, Theory of gelation, vitrification, and activated
  barrier hopping in mixtures of hard and sticky spheres.
\newblock {\em\protect\JournalTitle{J. Chem. Phys.}} \textbf{128}, 084509
  (2008).

\bibitem{Langmuir.28.1866-1878}
APR Eberle, R Casta{\~n}eda-Priego, JM Kim, NJ Wagner, Dynamical arrest,
  percolation, gelation, and glass formation in model nanoparticle dispersions
  with thermoreversible adhesive interactions.
\newblock {\em\protect\JournalTitle{Langmuir}} \textbf{28}, 1866--1878 (2012).

\bibitem{JFluidMech.852.P1}
{\'E} Guazzelli, O Pouliquen, Rheology of dense granular suspensions.
\newblock {\em\protect\JournalTitle{J. Fluid Mech.}} \textbf{852}, P1 (2018).

\bibitem{JCIS.598.56-68}
AJ Gravelle, AG Marangoni, A new fractal structural-mechanical theory of
  particle-filled colloidal networks with heterogeneous stress translation.
\newblock {\em\protect\JournalTitle{J. Colloid Interface Sci.}} \textbf{598},
  56--68 (2021).

\bibitem{PhysRevRes.4.043181}
S Zhang, et~al., Prestressed elasticity of amorphous solids.
\newblock {\em\protect\JournalTitle{Phys. Rev. Res.}} \textbf{4}, 043181
  (2022).

\bibitem{AdvColloidInterfaceSci.73.71-126}
W Poon, M Haw, Mesoscopic structure formation in colloidal aggregation and
  gelation.
\newblock {\em\protect\JournalTitle{Adv. Colloid Interface Sci.}} \textbf{73},
  71--126 (1997).

\bibitem{Macromolecules.51.5375-5391}
V Sorichetti, V Hugouvieux, W Kob, {Structure and Dynamics of a
  Polymer-Nanoparticle Composite: Effect of Nanoparticle Size and Volume
  Fraction}.
\newblock {\em\protect\JournalTitle{Macromolecules}} \textbf{51}, 5375--5391
  (2018).

\bibitem{ChemCommun.1186-1187}
B Tian, et~al., Microwave assisted template removal of siliceous porous
  materials.
\newblock {\em\protect\JournalTitle{Chem. Commun.}} pp. 1186--1187 (2002).

\bibitem{PhysRevA.42.4772}
WH Shih, WY Shih, SI Kim, J Liu, IA Aksay, Scaling behavior of the elastic
  properties of colloidal gels.
\newblock {\em\protect\JournalTitle{Phys. Rev. A}} \textbf{42}, 4772--4779
  (1990).

\bibitem{PhysRevLett.120.208005}
JM van Doorn, JE Verweij, J Sprakel, J van~der Gucht, Strand plasticity governs
  fatigue in colloidal gels.
\newblock {\em\protect\JournalTitle{Phys. Rev. Lett.}} \textbf{120}, 208005
  (2018).

\bibitem{JRheol.58.1089-1116}
J Colombo, E Del~Gado, Stress localization, stiffening, and yielding in a model
  colloidal gel.
\newblock {\em\protect\JournalTitle{J. Rheol.}} \textbf{58}, 1089--1116 (2014).

\bibitem{PNAS.121.e2316394121}
M Nabizadeh, et~al., Network physics of attractive colloidal gels: Resilience,
  rigidity, and phase diagram.
\newblock {\em\protect\JournalTitle{Proc. Natl. Acad. Sci. U.S.A.}}
  \textbf{121}, e2316394121 (2024).

\bibitem{JStatPhys.93.603-613}
NV Dokholyan, et~al., Scaling of the distribution of shortest paths in
  percolation.
\newblock {\em\protect\JournalTitle{J. Stat. Phys.}} \textbf{93}, 603--613
  (1998).

\bibitem{NatPhys.19.1171-1177}
H Tsurusawa, H Tanaka, Hierarchical amorphous ordering in colloidal gelation.
\newblock {\em\protect\JournalTitle{Nat. Phys.}} \textbf{19}, 1171--1177
  (2023).

\bibitem{ChemCommun.1998.2203-2204}
M T.~J.~Keene, R Denoyel, P L.~Llewellyn, Ozone treatment for the removal of
  surfactant to form mcm-41 type materials.
\newblock {\em\protect\JournalTitle{Chem. Commun.}} pp. 2203--2204 (1998).

\bibitem{EnergyFuels.36.2424-2446}
H Ghaedi, M Zhao, Review on template removal techniques for synthesis of
  mesoporous silica materials.
\newblock {\em\protect\JournalTitle{Energy Fuels}} \textbf{36}, 2424--2446
  (2022).

\bibitem{Biomater.25.3793-3805}
XZ Zhang, DQ Wu, CC Chu, Synthesis, characterization and controlled drug
  release of thermosensitive ipn--pnipaam hydrogels.
\newblock {\em\protect\JournalTitle{Biomater.}} \textbf{25}, 3793--3805 (2004).

\bibitem{IntJBioprint.8.599}
M Xie, et~al., Thermo-sensitive sacrificial microsphere-based bioink for
  centimeter-scale tissue with angiogenesis.
\newblock {\em\protect\JournalTitle{Int. J. Bioprint}} \textbf{8}, 599 (2022).

\bibitem{TFST.86.85-94}
T Farjami, A Madadlou, An overview on preparation of emulsion-filled gels and
  emulsion particulate gels.
\newblock {\em\protect\JournalTitle{Trends Food Sci. Technol.}} \textbf{86},
  85--94 (2019).

\bibitem{FoodHydrocoll.119.106875}
AJ Gravelle, AG Marangoni, Effect of matrix architecture on the elastic
  behavior of an emulsion-filled polymer gel.
\newblock {\em\protect\JournalTitle{Food Hydrocoll.}} \textbf{119}, 106875
  (2021).

\bibitem{SoftMatter.19.2771}
KW Torre, J de~Graaf, Structuring colloidal gels via micro-bubble oscillations.
\newblock {\em\protect\JournalTitle{Soft Matter}} \textbf{19}, 2771--2779
  (2023).

\bibitem{SoftMatter.18.2092}
B Saint-Michel, G Petekidis, V Garbin, Tuning local microstructure of colloidal
  gels by ultrasound-activated deformable inclusions.
\newblock {\em\protect\JournalTitle{Soft Matter}} \textbf{18}, 2092--2103
  (2022).

\bibitem{NatCommun.14.5309}
FJ M{\"u}ller, L Isa, J Vermant, Toughening colloidal gels using rough building
  blocks.
\newblock {\em\protect\JournalTitle{Nat. Commun.}} \textbf{14}, 5309 (2023).

\bibitem{ComputPhysCommun.271.108171}
AP Thompson, et~al., Lammps - a flexible simulation tool for particle-based
  materials modeling at the atomic, meso, and continuum scales.
\newblock {\em\protect\JournalTitle{Comput. Phys. Commun.}} \textbf{271},
  108171 (2022).

\bibitem{ModelSimulMaterSciEng.18.015012}
A Stukowski, Visualization and analysis of atomistic simulation data with
  ovito--the open visualization tool.
\newblock {\em\protect\JournalTitle{Model. Simul. Mater. Sci. Eng.}}
  \textbf{18}, 015012 (2009).

\bibitem{SoftMatter.15.4150423}
R Cabriolu, J Horbach, P Chaudhuri, K Martens, Precursors of fluidisation in
  the creep response of a soft glass.
\newblock {\em\protect\JournalTitle{Soft Matter}} \textbf{15}, 415--423 (2019).

\end{thebibliography}
\end{document}